# The Origin of Tunneling Anisotropic Magnetoresistance in Break Junctions

J. D. Burton,[1,3*] R. F. Sabirianov,[2,3] J. P. Velev,[1,3] O. N. Mryasov,[4] and E. Y. Tsymbal[1,3**]

[1]*Department of Physics and Astronomy, University of Nebraska, Lincoln, Nebraska 68588-0111, USA*
[2]*Department of Physics, University of Nebraska, Omaha, Nebraska 68182-0266, USA*
[3]*Nebraska Center for Materials and Nanoscience University of Nebraska, Lincoln, Nebraska 68588-0111, USA*
[4]*Seagate Research, Pittsburgh, Pennsylvania 15222, USA*

First-principles calculations of electron tunneling transport in Ni and Co break junctions reveal strong dependence of the conductance on the magnetization direction, an effect known as tunneling anisotropic magnetoresistance (TAMR). The origin of this phenomenon stems from resonant states localized in the electrodes near the junction break. The energy and broadening of these states is strongly affected by the magnetization orientation due to spin-orbit coupling, causing TAMR to be sensitive to bias voltage on a scale of a few mV. Our results bear a resemblance to recent experimental data and suggest that TAMR driven by resonant states is a general phenomenon typical for magnetic broken contacts and other experimental geometries where a magnetic tip is used to probe electron transport.

Anisotropic magnetoresistance (AMR) is the difference in resistance of a magnetic conductor as the magnetization direction is changed with respect to the direction of current flow. In bulk ferromagnets this phenomenon originates from the anisotropy of scattering produced by spin-orbit coupling (SOC) in the diffusive transport regime.[1] The nature of AMR in nanoscale magnetic systems such as ballistic conductors or tunnel junctions is profoundly different due to different mechanisms that control electron transport. For example, in magnetic nanocontacts the ballistic nature of electron transport leads to quantized conductance as the angle of the magnetization is altered with respect to the axis of the wire.[2,3] This phenomenon is known as ballistic anisotropic magnetoresistance (BAMR). Another kind of AMR is inherent to the tunneling transport regime and is called tunneling anisotropic magnetoresistance (TAMR). TAMR has been observed in tunnel junctions with dilute magnetic semiconductor electrodes.[4] It was also detected by scanning tunneling spectroscopy of thin Fe films on W(110) substrates[5] and predicted for tunneling from metallic alloys with large magnetocrystalline anisotropy, such as CoPt,[6] and from the Fe(100) surface through vacuum.[7] TAMR originates from the anisotropy of the electronic structure induced by SOC. It is different from the usual tunneling magnetoresistance (TMR).[8] While TMR is due to the reorientation of magnetic moments of the two magnetic electrodes relative to one another, TAMR is intrinsic to magnetically saturated tunnel junctions and may occur in junctions with only one magnetic electrode.

Recently Bolotin *et al.*[9] have shown that atomic scale permalloy break junctions exhibit TAMR. By measuring the conductance of the completely broken contacts as a function of magnetic-field direction at saturation Bolotin *et al.* found that the TAMR can be as large as 25%. They also found that magnetoresistance was sensitive to changes in bias on a scale of a few mV. The origin of these phenomena was attributed to conductance fluctuations due to quantum interference effects based on a model proposed by Adam *et al.*[10] This model was, however, developed for *diffusive metallic* samples and, though it may be relevant to unbroken metallic junctions, it may not be applicable to tunneling conduction across broken contacts. Therefore, the origin of the large TAMR observed by Bolotin *et al.* and the source of its sensitivity to applied bias voltage remain elusive.

In this Letter, we perform first-principles calculations of the electronic structure and conductance of Ni and Co break junctions, where electron transport occurs via tunneling. We find a strong dependence of the tunneling conductance on the saturation magnetization direction, signature of the TAMR effect. We demonstrate that the TAMR is controlled by resonant states localized at the electrode tips near the break, which we refer to as *tip resonant states*. The energy and broadening of these states depend strongly on the magnetization orientation due to the SOC. Our results at finite bias show sensitivity of TAMR on a scale of a few mV, and the angular dependence of TAMR bears a striking resemblance to the experimental results of Bolotin *et al.*,[9] clearly indicating the origin of the observed phenomenon. We infer that TAMR driven by tip resonant states is a general phenomenon typical for magnetic broken contacts and other experimental geometries where a magnetic tip is used to probe electron transport.

We consider Ni break junctions consisting of two free standing semi-infinite nanowire electrodes made of ferromagnetic fcc Ni which are separated by a vacuum region, as shown in Fig. 1. The nanowires are built along the [001] direction ($z$ axis) by periodic repetition of a supercell made up of two topologically different (001) planes, one consisting of 5 atoms and the other of 4 atoms. We use the lattice constant $a = 3.52$ Å of bulk fcc Ni. The tip of each electrode has one apex atom and the separation between the two apex atoms is equal to three atomic planes, or 5.28Å.

Self-consistent density functional calculations of the spin-dependent electronic structure of the Ni break junctions are performed using the real space recursion method[11] with a tight-binding linear muffin-tin-orbital (TB-LMTO) basis[12] in the atomic sphere approximation.[13] The local spin density approximation is used for the exchange-correlation energy.

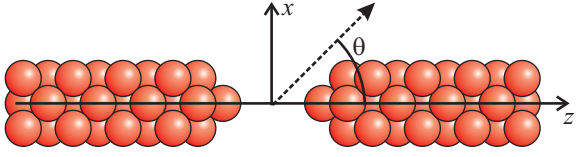

FIG. 1 (color online) The atomic structure of the break junction. The $z$ axis is the axis of the wire. The angle $\theta$ denotes the direction of the spin magnetic moments in the $xz$ plane.

The SOC is included within a scalar relativistic approximation via perturbation for each atomic sphere. We assume that all atomic spheres have the same spin orientation, and the magnetization lies in the $xz$ plane making angle $\theta$ with respect to the $z$ axis (Fig. 1). Therefore, the SOC term in the Hamiltonian is

$$\lambda \mathbf{L} \cdot \mathbf{S} = \frac{1}{2}\lambda \big[ L_x (\sigma_z \sin\theta + \sigma_x \cos\theta) + L_y \sigma_y \\ + L_z (\sigma_z \cos\theta - \sigma_x \sin\theta) \big] \quad (1)$$

where $\sigma$ is the Pauli spin matrix, $\mathbf{L}$ and $\mathbf{S}$ are the orbital and spin momentum operators, and $\lambda$ is the SOC constant.

To calculate transport properties of this break junction we appeal to the Landauer-Büttiker formalism of quantum transport[14] where the electric current is given by

$$I = \frac{e}{h}\int_{-\infty}^{\infty} T(E)\big[f(E-\mu_L) - f(E-\mu_R)\big]dE. \quad (2)$$

Here $T(E)$ is the transmission probability for states with energy $E$, $f(E)$ is the Fermi distribution function and $\mu_L$ ($\mu_R$) is the electrochemical potential of the left (right) electrode.[15] The transmission function is given by

$$T(E) = Tr\big[(\Sigma_L^\dagger - \Sigma_L) G(E)(\Sigma_R - \Sigma_R^\dagger) G^\dagger(E)\big], \quad (3)$$

where $G(E)$ is the total Green's function, and $\Sigma_L$ and $\Sigma_R$ are the self energies of the left and right electrodes respectively. In the absence of applied bias, tight-binding techniques implemented within the TB-LMTO method can be used to calculate $T(E)$ directly from Eq. (3),[16] where the necessary real-space Hamiltonian blocks are taken from the fully self-consistent calculation.

To evaluate the dependence on bias we use a simplified description, in the spirit of the Bardeen approximation.[17] We assume that the coupling through vacuum is weak so that each electrode can be treated as an independent sub-system. We consider a tight-binding layer within the barrier (which, in our case, is the vacuum layer of empty spheres in the plane $z = 0$) and assume that the left and right electrodes in Eq. (3) include the metal electrodes and a portion of the barrier up to that layer. If the layer lies in the barrier at a sufficient distance from both metal surfaces, the self-energies $\Sigma_L$ and $\Sigma_R$ entering Eq. (3) can be treated as small. Retaining only terms of the order of $\Sigma^2$ in the expression for the transmission (3), we can replace the Green's function $G(E)$ of the coupled system evaluated at the layer by the Green's function of the uncoupled barrier $G_0(E)$. Within the same approximation we can write the expression for the tunneling density of states (DOS) operator in the barrier induced by the left and right electrodes, $\rho_L$ and $\rho_R$ respectively, as follows

$$2\pi i \rho_{L,R}(E) = G_0(E)\big(\Sigma_{L,R} - \Sigma_{L,R}^\dagger\big) G_0(E). \quad (4)$$

This allows us to re-write (3) as

$$T(E) = 4\pi^2 Tr\big[G_0^{-1}(E)\rho_L(E) G_0^{-1}(E)\rho_R(E)\big], \quad (5)$$

where the trace is taken over all the orbital and spin indices over sites lying within the tight-binding layer in the barrier. The transmission has been previously derived directly from the Bardeen approximation using the embedding Green's function approach.[18]

If only one orbital state dominates the conductance, which is typical for a not too thin vacuum barrier,[19] the expression for the transmission can further be simplified. Neglecting non-diagonal components of the DOS operator and the Green's function,[20] we find

$$T(E) \approx 4\pi^2 \phi^2(E)\big[\rho_L^\uparrow(E)\rho_R^\uparrow(E) + \rho_L^\downarrow(E)\rho_R^\downarrow(E)\big], \quad (6)$$

where $\rho_{L,R}^\sigma(E)$ is the tunneling DOS for spin projection $\sigma$ along the magnetization axis, and $\phi(E)$ is the diagonal component of $G_0^{-1}(E)$ which has the meaning of an effective barrier potential seen by tunneling electrons. Eq. (6) is reminiscent of the approximation used to derive Jullière's formula,[21] however there is an important difference. According to Eq. (6) the transmission is determined by the tunneling DOS in the *barrier*, while Jullière's formula is expressed in terms of the DOS of the *electrodes*.

In our case the break junction is symmetric and the tunneling DOS induced by the left and right electrodes are $\rho_L^\sigma(E) = \tfrac{1}{2}\rho^\sigma(E-eV)$ and $\rho_R^\sigma(E) = \tfrac{1}{2}\rho^\sigma(E)$, where $V$ is an applied bias voltage and $\rho^\sigma(E)$ is the tunneling DOS induced by the *two* electrodes at *zero* bias, which can be taken directly from the band structure calculation. For small $V$ of a few mV we can neglect the variation of $\phi(E)$ with energy in Eq. (6). We verify the assumptions leading to Eq. (6) by calculating $T(E)$ using both Eqs. (3) and (6) in the absence of applied bias. The calculations reveal excellent agreement between the two formulas for all orientations of the magnetization when choosing an effective barrier height $\phi \approx 1.98$ eV (see Fig. 2a,d and g).[22]

Fig. 2a shows the transmission function calculated in the absence of SOC within ±30meV of the Fermi energy ($E_F$). Comparing Fig. 2a to the band structure of the Ni electrode shown in Fig. 2b, we see that at $E_F$ the tunneling transmission is dominated by states on the apex atoms originating from a minority spin $d$ band (marked $A$ in Fig.

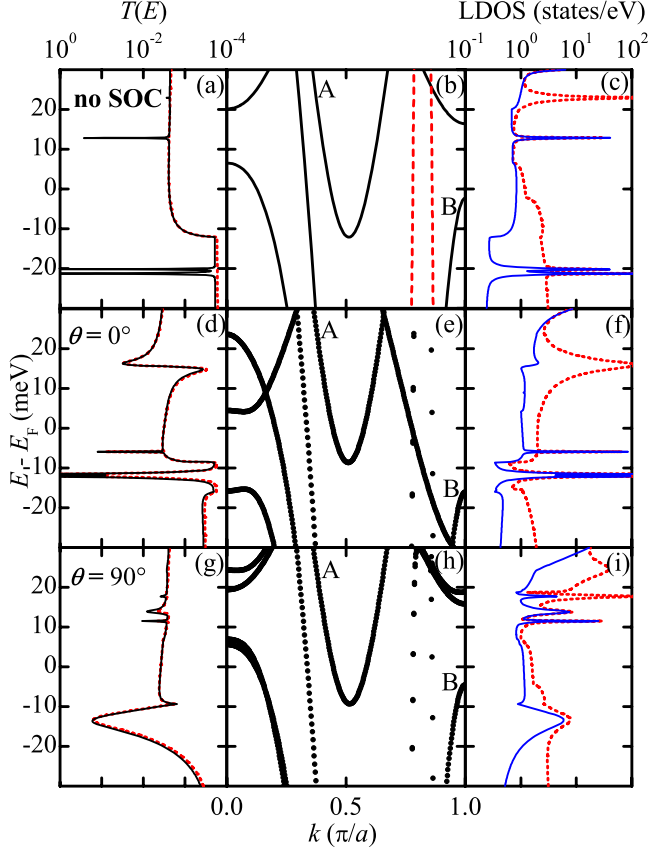

FIG. 2 (color online) The transmission function, band structure of the electrodes, and the local DOS on the electrode tips without SOC (a-c) and with SOC for $\theta = 0°$ (d-f) and $\theta = 90°$ (g-i). In (b) the solid curves are minority-spin bands, and the dashed curves are majority-spin bands. In (a), (d) and (g) the solid curves are calculated using Eq. (6) and the dashed curves are calculated using Eq. (3). In (c), (f) and (i) the solid curves are the minority DOS on the apex atom and the dashed curves are the minority DOS on the first sub-apex atomic layer.

2b). This is evident from the distinctive drop at the bottom of the band in the local DOS at the apex atom (Fig. 2c) and in the transmission function (Fig. 2a). Below the band minimum at −20 meV, there appear two narrow peaks in the transmission due to tip resonant states (Fig. 2a).[23] These resonant states are seen in Fig. 2c and arise because the lower coordination of the apex atom reduces the effective onsite energy so that the level lies outside of the continuum of band $A$.[24] These tip resonant states have minority $d_{z^2}$ character, similar to the states of band $A$. The weak interaction through vacuum splits these states into a bonding/antibonding pair.[25] Another tip resonant state at 12.8 meV (Figs. 2a,c) is double degenerate and is composed of the $d_{xz}$ and $d_{yz}$ orbitals. One more tip resonant state barely visible in Fig. 2a at 23 meV is localized on the four atoms in the first sub-apex atomic layer (the dashed line in Fig. 2c).

SOC makes the tip resonant states strongly dependent on the magnetization angle $\theta$. The most important effect for the TAMR is the broadening of the $d_{z^2}$ tip resonant state (the bottommost resonance in Figs. 2d,f and 2g,i). The broadening of this state is smaller for $\theta = 0°$ because the SOC only mixes minority $d_{z^2}$ states with majority $d$ states, and not with other minority $d$ states.[26] However, when $\theta = 90°$ the SOC (1) mixes the minority $d_{z^2}$ states with minority $d_{xz}$ states resulting in stronger broadening. In addition, the minority $d$ band that has a maximum just below $E_F$ (marked $B$ in Figs. 2b,e,h) at the Brillouin zone edge has significant $d_{xz}$ character and the tip resonance can hybridize with this band, provided this band overlaps with the resonance level. Since the top of this band also varies with $\theta$, only a certain range of $\theta$ will meet this condition, leading to the transition from the sharp double-peak feature of Figs. 2d,f to the broad resonance in Figs. 2g,i.

Another effect of the SOC on the transmission function is the splitting of the $d_{xz},d_{yz}$ doublet (the peak at 12.8 meV in Figs. 2a,c) into two singlets. The splitting is largely determined by the $\frac{1}{2}\lambda L_z \sigma_z \cos\theta$ term in Eq. (1), making the splitting much stronger for $\theta = 0°$ than for $\theta = 90°$. For $\theta = 0°$ the lower state of the split doublet is seen in Figs. 2d,f as a sharp peak at −6 meV. The upper state lies outside of the energy range plotted. For $\theta = 90°$ the splitting is much weaker and occurs due to the coupling of the doublet states via other states (e.g., the bulk band $A$), producing two peaks at 11.8 meV and 17.4 meV, as seen in Figs. 2g,i.

One more feature induced by the SOC is the Fano-shaped resonance[27] seen in Fig. 2d at around 15 meV for $\theta = 0°$. It mirrors a related feature that appears in the local DOS on the apex atom (Fig. 2f). This resonance is due to the localized state that originates on the first sub-apex layer of atoms. In the absence of SOC this state, appearing at 23 meV in Fig. 2c, is not coupled to the band $A$ and does not appear in the DOS on the apex atom. However, when SOC is included, the state becomes coupled to the band $A$ and, by propagating through the band to the apex atom, acquires the Fano shape. For $\theta = 90°$ the Fano resonance, appearing at 14 meV in Fig. 2i, becomes less pronounced due to a weaker coupling between the resonant and continuum states.

The effect of bias is demonstrated in Fig. 3. Here we plot the differential conductance $G(V) = dI/dV$ and its angular dependence. The current $I(V)$ is calculated using Eqs. (2) and (6), assuming the Fermi distribution function at 4.2 K. As is evident from Fig. 3a, the angular dependence of the conductance reflecting TAMR is very different from the $\cos^2\theta$ dependence typical for bulk AMR.[1] The magnitude of TAMR is relatively large, e.g. it reaches 200% for $V = 11.8$ mV. The conductance variation is quite sensitive to the applied bias, owing to the sensitivity of $\rho^\sigma(E)$ to $\theta$ due to resonant states. Fig. 3b shows the bias dependence for a few values of $\theta$. The central peak in Fig. 3b is due to the $d_{z^2}$ tip resonance state which broadens with increasing angle from 0° to 90°. The Fano resonance can be seen between 14 and 16 mV for $\theta = 0°$. Although finite temperature and bias smear out the sharp resonant features seen in Figs. 2d,g, the angular dependence of the conductance is largely controlled

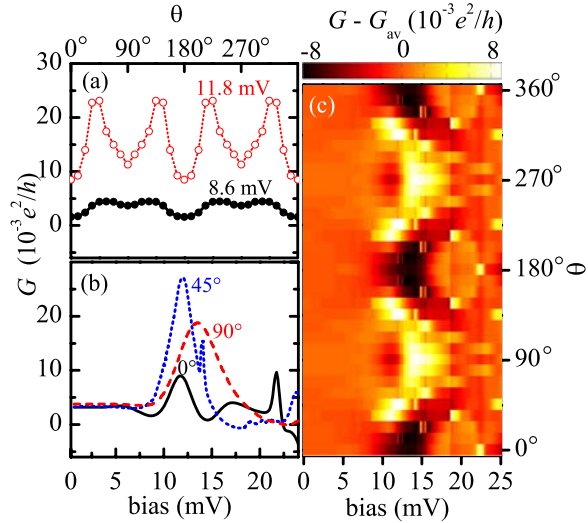

FIG. 3 (color online) Variations of differential conductance $G(V) = dI/dV$ at 4.2K. (a) $G$ verses $\theta$ for different bias voltages. (b) Dependence of $G$ on bias for different $\theta$. (c) Deviation of $G$ from $G_{av}$, the conductance averaged over $\theta$, as a function of $V$ and $\theta$.

by the tip resonant states. Fig. 3c shows the deviation of $G(V)$ from the value $G_{av}(V)$, the average over $\theta$ for applied bias $V$. Clearly the deviation is strongly dependent on the bias voltage, and the central light/dark pattern is due to the variation of the $d_{z^2}$ tip resonant state with $\theta$. The resemblance of Fig. 3c to the plot of Fig. 4d of Bolotin et al.[9] shows that tip resonance states play a decisive role in TAMR.

The importance of tip resonant states is further supported by calculations performed for another geometry and material. For Ni junctions having larger radius,[28] we find that tip resonant states lead to a very similar bias dependent TAMR. This indicates that the effect is most sensitive to the features of the tip, and to a lesser extent the radius of the electrodes. For a Co junction several tip resonant states are present, however at different energies than those for Ni. In particular, a $d_{z^2}$ tip resonance lies at 50 meV above the Fermi energy and exhibits similar variation with $\theta$ as the one shown in Fig. 3c. These results clearly demonstrate that TAMR driven by resonant states is a general phenomenon intrinsic to magnetic broken contacts.

In conclusion, we would like to emphasize the fact that the predicted phenomenon may occur in any geometry where a magnetic tip is used to probe electron transport. In particular, the effect may be important and studied using spin-polarized scanning-tunneling spectroscopy,[5] where a magnetic tip may be used to scan a nonmagnetic sample. We hope, therefore, that this Letter will stimulate further experimental studies of the influence of tip resonant states on TAMR.

This work is supported by Seagate Research and the NSF-MRSEC (Grant No. DMR-0213808). We thank Dan Ralph and Kirill Belashchenko for helpful discussions.

* Electronic address: jlz101@unlserve.unl.edu
** Electronic address: tsymbal@unl.edu